\documentclass[aps,pra,showpacs,twocolumn, longbibliography]{revtex4-1}

\usepackage{graphicx}
\usepackage{dcolumn}
\usepackage{bm}
\usepackage{color}
\usepackage{amsmath}
\usepackage{amssymb}

\newcommand{\tabincell}[2]{\begin{tabular}{@{}#1@{}}#2\end{tabular}}

\begin{document}

\title{Spectrally uncorrelated biphotons generated from `the family of BBO crystal' }

\author{Rui-Bo Jin$^{1,3}$}
\author{Wu-Hao Cai$^{1}$}
\author{Chunling Ding$^{1}$}
\author{Feng Mei$^{3, 4}$}
\email{meifeng@sxu.edu.cn}
\author{Guang-Wei Deng$^{2,6}$}
\email{gwdeng@uestc.edu.cn}
\author{Ryosuke Shimizu$^{5}$}
\author{Qiang Zhou$^{2, 6}$}
\email{zhouqiang@uestc.edu.cn}

\affiliation{$^{1}$ Hubei Key Laboratory of Optical Information and  Pattern Recognition, Wuhan Institute of Technology, Wuhan 430205, China}
\affiliation{$^{2}$ Institute of Fundamental and Frontier Science and School of Optoelectronic Science and Engineering, University of Electronic Science and Technology of China, Chengdu 610054, China}
\affiliation{$^{3}$ State Key Laboratory of Quantum Optics and Quantum Optics Devices, Institute of Laser Spectroscopy, Shanxi University, Taiyuan, Shanxi 030006, China}
\affiliation{$^{4}$ Collaborative Innovation Center of Extreme Optics, Shanxi University, Taiyuan, Shanxi 030006, China}
\affiliation{$^{5}$The University of Electro-Communications, 1-5-1 Chofugaoka, Chofu, Tokyo, Japan}
\affiliation{$^{6}$CAS Key Laboratory of Quantum Information, University of Science and Technology of China, Hefei 230026, China}
\date{\today }

\begin{abstract}
Spectrally intrinsically uncorrelated biphoton states generated from nonlinear crystals are very important but rare resources for quantum photonics and quantum information applications.
Previously, such biphoton states were generated from several kinds of crystals, however, their wavelength ranges and nonlinear efficiencies were still limited for various applications.
In order to explore new crystal for wider wavelength range and higher nonlinear efficiency, here we theoretically study the generation of spectrally uncorrelated biphoton states from 14 crystals in the `BBO family', including BBO, CLBO, KABO, KBBF, RBBF, CBBF, BABF, BiBO, LBO, CBO, LRB4, LCB, YCOB, and GdCOB.  They satisfy three kinds of group-velocity matching condition from near-infrared to telecom wavelengths.
Furthermore, heralded single photons can be generated with a purity as high as 0.98, which is achieved without any narrow filtering. The indistinguishability of photons from independent sources is examined by the Hong-Ou-Mandel interference, which results in a visibility of  98\% also without any further filtering, i.e., photons from different heralded single-photon sources are highly indistinguishable. Our study may provide single-photon sources with good performance for quantum information processing at near-infrared and telecom wavelengths.
\end{abstract}



\maketitle

\section{Introduction}
Biphotons generated from a spontaneous parametric down-conversion (SPDC) process in a nonlinear crystal play an important role in photonic quantum information processing (QIP) \cite{Slussarenko2019APR, Flamini2018RPP, Couteau2018}.
Thanks to the energy and momentum conservation laws, the biphotons generated are generally spectrally correlated,  which leads to low collection efficiency and unwanted noise photons for further application, for instance in the application of heralded single photon generation. An avenue to improve the collection efficiency and signal to noise ratio is removing the spectral correlation of generated biphoton state, i.e., spectrally uncorrelated biphotons.
Recently, such biphoton states and the heralded single photons from them are on demand for many applications in QIP, for example, quantum computation \cite{Walmsley2005}, boson sampling \cite{Broome2013}, quantum teleportation \cite{Valivarthi2016np} and measurement-device-independent quantum key distribution (MDIQKD) \cite{Lo2012, Zhou2017QST}.
Two methods have been demonstrated to remove the spectral correlations. One is spectral filtering with narrow bandpass filters, which is simple but has drawbacks of severely decreasing the brightness and  thus lowing the heralding efficiency of the heralded photon sources \cite{Meyer-Scott2017}.  The second method is to design the nonlinear crystal, so as to generate spectrally uncorrelated biphotons through group-velocity-matched (GVM) conditions \cite{Grice2001,  Konig2004, Edamatsu2011}.

\begin{table*}[!bhtbp]
\centering\begin{tabular}{c|cccccc}
\hline \hline
Name &chemical formula         &axis      &structure unit   &point group    & $\lambda_{transp.}$ (nm)  & $d_{max}$ (pm/V)  \\
 \hline
BBO &$\mathrm{\beta-BaB_2O_4}$ & uniaxial  &$\mathrm{ (B_3 O_6)^{3-} }$  &3m  &  189$\sim$3500 &$d_{22}$=2.2 \\
 \hline
CLBO &$\mathrm{CsLiB_6O_{10}}$ & uniaxial  &$\mathrm{ (B_3O_7)^{5-} }$  & $\bar4$2m & 180$\sim$2750 &$d_{36}$=0.74 \\
 \hline
KABO &$\mathrm{K_2Al_2B_2O_7}$ & uniaxial  &$\mathrm{ (BO_3)^{3-} }$  &32   & 180$\sim$3600 &$d_{11}$=0.45 \\
 \hline
KBBF &$\mathrm{KBe_2BO_3F_2}$ & uniaxial  &$\mathrm{ (BO_3)^{3-} }$  &32   & 155$\sim$3700 &$d_{11}$=0.49 \\
 \hline
RBBF &$\mathrm{RbBe_2BO_3F_2}$ & uniaxial  &$\mathrm{ (BO_3)^{3-} }$  &32   & 160$\sim$3550 &$d_{11}$=0.45 \\
 \hline
CBBF &$\mathrm{CsBe_2BO_3F_2}$ & uniaxial  &$\mathrm{ (BO_3)^{3-} }$  &32   & 151$\sim$3700 &$d_{11}$=0.5 \\
 \hline
BABF &$\mathrm{BaAlBO_3F_2}$ & uniaxial  &$\mathrm{ (BO_3)^{3-} }$  &$\bar6$   & 165$\sim$3000 &$d_{22}$=1.24 \\
 \hline
 \hline
BiBO &$\mathrm{BiB_3O_6}$ & biaxial  &$\mathrm{ (BO_3)^{3-}, (BO_4)^{5-}}$  &2   & 270$\sim$2730 &$d_{26}$=3.48 \\
 \hline
LBO &$\mathrm{LiB_3O_5}$ & biaxial  &$\mathrm{ (B_3O_7)^{5-} }$  &mm2   & 155$\sim$2600 &$d_{32}$=0.85 \\
\hline
CBO &$\mathrm{CsB_3O_5}$ & biaxial  &$\mathrm{ (B_3O_7)^{5-} }$  &222   & 170$\sim$3000 &$d_{14}$=1.08 \\
 \hline
LRB4 &$\mathrm{LiRbB_4O_7}$ & biaxial  &$\mathrm{ (B_3O_7)^{5-} }$  &222   & 187$\sim$3468 &$d_{14}$=0.45 \\
 \hline
LCB &$\mathrm{La_2CaB_{10}O_{19}}$ & biaxial  &$\mathrm{ (B_5O_{12})^{9-}}$  &2   & 185$\sim$3000 &$d_{22}$=-1.04 \\
 \hline
YCOB &$\mathrm{YCa_4O(BO_3)_3}$ & biaxial  &$\mathrm{ (BO_{3})^{3-}}$  &m   & 202$\sim$3700 &$d_{22}$=2.03 \\
 \hline
GdCOB &$\mathrm{GdCa_4O(BO_3)_3}$ & biaxial  &$\mathrm{ (BO_{3})^{3-}}$  &m   & 200$\sim$3700 &$d_{22}$=2.23 \\
\hline \hline
\end{tabular}
\caption{\label{table:CrystalSummary} Main properties of the 14 borate crystals discussed in this work, including the chemical formula, the axis (uniaxial or biaxial), the basic structure unit, the point group, the transparency range $\lambda_{transp.}$, and  the maximal nonlinear coefficient $d_{max}$ at around 1064 nm wavelength. All the data was obtained from Refs. \cite{Chen2012BOOK, Dmitriev1999, Nikogosyan2005}.  }
\end{table*}

Several works have been demonstrated for spectrally uncorrelated biphotons generation using different GVM crystals at different wavelengths. For example, the KDP crystal at 830 nm has a maximal purity of 0.97 \cite{Mosley2008PRL, Jin2011, Jin2013PRA2};
11 isomorphs of KDP can also maintain high spectral purity at near-infrared and telecom wavelengths \cite{Jin2019PRAppl};
the  $\beta$-barium borate (BBO) crystal at 1526 nm can achieve a maximal purity of 0.82 \cite{Grice2001, Lutz2013OL, Lutz2014, TingyuLi2018OE}; the periodically poled KTP crystal (PPKTP) at 1584 nm has a maximal purity of 0.82 \cite{Evans2010, Gerrits2011, Eckstein2011, Jin2013OE, Liyan2015, Jin2018Optica, Greganti2018, Meyer-Scott2018OE, Terashima2018, Zhang2018PRA}, and this high purity  can be kept when the wavelength is tuned from 1400 nm to 1700 nm \cite{Jin2013OE}; the purity for PPKTP crystals can be further improved from 0.82 to near 1 using the custom poling crystal \cite{Branczyk2011, Dixon2013, Dosseva2016, Tambasco2016, Chen2017, Graffitti2018PRA, Graffitti2018Optica, Chen2019OE}, or even using a machine-learning framework \cite{Cui2019PRAppl};  4 isomorphs of PPKTP can retain the properties of PPKTP, i.e, these isomorphs satisfy the GVM condition and can prepare spectrally uncorrelated biphoton state in the range of 1300 nm to 2100 nm \cite{Jin2016PRAppl, Laudenbach2017, Jin2019JOLT}.

Generally speaking, each nonlinear crystal has two specific GVM wavelengths to achieve high purity of around 0.97 \cite{Jin2019PRAppl}, but the tunable range of those specific wavelengths is usually very narrow. In other words, these spectrally pure states can only be prepared at some wavelength points, not at the full wavelength line.  The custom poling technique can maintain both high purity and wide wavelength range,  but is only applied on PPKTP and limited in the telecom wavelength range, i.e, cannot cover the near-infrared or visible range \cite{Branczyk2011, Dixon2013, Dosseva2016, Tambasco2016, Chen2017, Graffitti2018PRA, Graffitti2018Optica, Chen2019OE}. The isomorphs of KDP increased the GVM wavelength range but with lower nonlinear efficiency \cite{Jin2019PRAppl}. Therefore, the generation of spectrally pure biphoton states from more nonlinear crystals with wider wavelength range and higher nonlinear efficiency is still a field with high demand.  In this work, we investigate the possibility of  generating spectrally uncorrelated biphoton state from 14 isomorphs from the `BBO family', i.e., the borate crystals, which provide wider GVM wavelengths and higher nonlinear efficiency.

\section{Theory }
\subsection{ Characteristics of the `BBO family'}
The widely used  borate crystals  include BBO, LBO and BiBO, which were discovered in 1978, 1987, and 1995 respectively \cite{Chen2012BOOK, Dmitriev1999, Nikogosyan2005}. Besides these three crystals, there are still many other borate  nonlinear optical crystals. Table\,\ref{table:CrystalSummary} summarized 14 typical  crystals from the `BBO family', including 7 uniaxial and 7 biaxial crystals \cite{Chen2012BOOK, Dmitriev1999, Nikogosyan2005}.
All these 7 uniaxial crystals are negative crystals with $n_e < n_o$ and the 7 biaxial crystals satisfy  $n_x < n_y < n_z$. Here $n_{e(o)}$ is the refractive index for extraordinary (ordinary) ray, and $n_{x(y,z)}$ is the refractive index for $X$-($Y$-, $Z$-) dielectric axis.
They mainly have  basic structure units of (B$_3$O$_6$)$^{3-}$, (B$_3$O$_7$)$^{5-}$,  (BO$_3$)$^{3-}$, (BO$_4$)$^{5-}$ and  (B$_5$O$_{12})^{9-}$. The point group, an important parameter determining the effective nonlinear coefficient, is including 3m, $\bar4$2m, 32, and $\bar6$ for the uniaxial crystals,  and 2, mm2, 222, and m for the biaxial crystal.  They have a wide transparent range (from UV to infrared), relative high nonlinear coefficient (up to 3.48 pm/V),  high damage threshold. These crystals have important applications, particularly in the laser industry.

In the quantum information field, BBO and BiBO are well-known crystals from the `BBO family'. BBO crystal is widely used to generate entangled photons with high brightness \cite{Kwiat1995, Kwiat1999}, especially in the case of beam-like configurations  \cite{Takeuchi2001, Niu2008, Xu2018}. Up to now, it has been used to prepare 6, 8, 10, 12 entangled photons \cite{Lu2007, Huang2011, Yao2012,  Zhonghansen2018}. BiBO has higher nonlinear coefficients and a smaller spatial walk-off angle than the BBO crystal. BiBO has also been used in many quantum optical experiments, e.g., quantum-enhanced tomography \cite{Zhou2015Optica}, 12 photon entangled state generation\cite{Chenluokan2017}.
In this work, we focus on the GVM conditions of these crystals, and use these crystals for spectrally uncorrelated biphoton state generation.

\subsection{The principle of spectrally uncorrelated biphoton state generation based on GVM conditions}
In the process of spontaneous parametric down-conversion (SPDC), a pump photon is  probabilistically converted to a biphoton state, e.g., the signal and idler, and in this process the energy and momentum is conserved.
The biphoton state $\vert\psi\rangle$ can be written as
\begin{equation}\label{eq1}
\vert\psi\rangle=\int_0^\infty\int_0^\infty\,\mathrm{d}\omega_s\,\mathrm{d}\omega_if(\omega_s,\omega_i)\hat a_s^\dag(\omega_s)\hat a_i^\dag(\omega_i)\vert0\rangle\vert0\rangle,
\end{equation}
where $\omega$ is the angular frequency; the subscripts $s$ and $i$ indicate the signal and idler photon respectively; $\hat a^\dag$ is the creation operator.
The joint spectral amplitude (JSA) $f(\omega_s,\omega_i)$ is the product of the pump-envelope function (PEF) $\alpha(\omega_s,\omega_i)$ and the phase-matching function (PMF) $\phi(\omega_s,\omega_i)$, i.e.,
\begin{equation}\label{eq2}
f(\omega_s,\omega_i) = \alpha(\omega_s,\omega_i) \times \phi(\omega_s,\omega_i).
\end{equation}
The PEF is determined by the energy conservation law, and for a Gaussian-distribution, PEF can be written as \cite{Mosley2008NJP}
\begin{equation}\label{eq20}
\alpha(\omega_s, \omega_i)=\exp[-\frac{1}{2}\left(\frac{\omega_s+\omega_i-\omega_{p_0}}{\sigma_p}\right)^2],
\end{equation}
where $\omega_{p_0}$ is the center wavelength of the pump;   $\sigma_p$ is the bandwidth of the pump, and the full-width at half-maximum (FWHM) is FWHM$_\omega$=2$\sqrt{\ln(2)} \sigma_p \approx 1.67\sigma_p$.

Using wavelengths as the variables, the PEF can be rewritten as
\begin{equation}\label{eq21}
\alpha(\lambda_s, \lambda_i)= \exp \left(  -\frac{1}{2} \left\{  \frac{{1/\lambda _s  + 1/\lambda _i  - 1/(\lambda _0 /2)}}{{\Delta \lambda /[(\lambda _0 /2)^2  - (\Delta \lambda /2)^2 ]}}  \right\}^2 \right),
\end{equation}
where  $\lambda _0 /2 $ is the central wavelength of the pump; The FWHM of the pump at intensity level is
FWHM$_\lambda$=$ \frac{2\sqrt{\ln (2)} {\lambda _0}^2   \Delta \lambda    \left({\lambda _0}^2-\Delta \lambda ^2\right)}{{\lambda _0}^4+\Delta \lambda ^4-2 {\lambda _0}^2 \Delta \lambda ^2 [1+\ln (4)]} $.
For $\Delta \lambda << \lambda _0$, FWHM$_\lambda\approx 2\sqrt{\ln(2)} \Delta \lambda  \approx 1.67\Delta \lambda $.

\begin{figure}[tbp]
\centering\includegraphics[width=8cm]{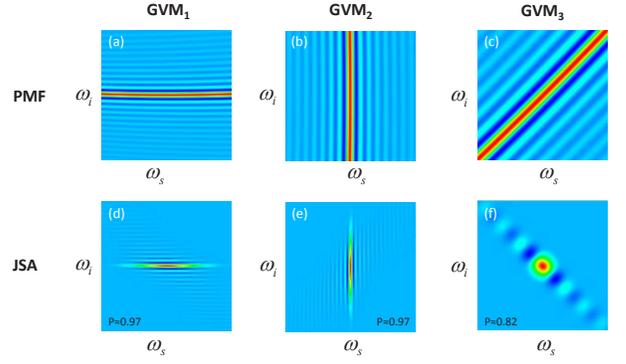}
\caption{ The concept of three group-velocity matching (GVM) conditions. Figures in the first row show the phase-matching function (PMF) distributed along the vertical, horizontal and diagonal direction for GVM$_1$,  GVM$_2$, and GVM$_3$, respectively. The second row shows the corresponding joint spectral amplitude (JSA) with the maximal purity of around 0.97, 0.97, and 0.82 respectively.
 } \label{Fig:concept}
\end{figure}

The PMF function is determined by the momentum conservation law.
By assuming a flat phase distribution, the PMF function can be written as \cite{Mosley2008NJP}
\begin{equation}\label{eq22}
\phi(\omega_s,\omega_i)=\operatorname{sinc}\left(\frac{\Delta kL}{2}\right),
\end{equation}
where $L$ is the length of crystal;  $\Delta k=k_p-k_i-k_s$ and $k=\frac{2 \pi n (\lambda, \theta, \varphi)}{\lambda}$ is the wave vector; $n$ is the refractive index;  $\theta$ and $\varphi$ are the phase-matching angles; $\theta$ is the polar angle and $\varphi$ is the azimuth angle,  as indicated in the spherical  coordinate in the Appendix.
According to the theoretical analysis, the shape of the PEF is determined by the GVM condition \cite{Grice2001, Edamatsu2011, Jin2013OE}.
In this work, we study three kinds of GVM conditions as following \cite{Jin2019PRAppl},
\begin{equation} \label{Equ:gvm}
\begin{array}{ll}
  GVM_1:  &  V_{g,p}^{-1}(\omega_p)=V_{g,s}^{-1}(\omega_s), \\
  GVM_2:  & V_{g,p}^{-1}(\omega_p)=V_{g,i}^{-1}(\omega_i), \\
  GVM_3:  & 2V_{g,p}^{-1}(\omega_p)=V_{g,i}^{-1}(\omega_i)+V_{g,s}^{-1}(\omega_s).
\end{array}
\end{equation}
Here  $V_{g,\mu}=\frac{d\omega}{dk_\mu(\omega)}=\frac{1}{k_\mu^\prime(\omega)},(\mu=p, s, i)$ is the group velocity of the pump, the signal and the idler.
Under the GVM$_1$, GVM$_2$ or GVM$_3$ conditions, the corresponding PMFs are distributed along the horizontal, vertical, and diagonal directions, as shown in Fig.\,\ref{Fig:concept}(a-c). With these three kinds of GVM conditions, it is possible to prepare spectrally uncorrelated biphotons and spectrally pure heralded single photons.
The JSAs with the maximal purity under these three GVM conditions are shown in Fig.\,\ref{Fig:concept}(d-f). The corresponding maximal purity is around 0.97, 0.97, and 0.82.
The purity can be calculated using Schmidt decomposition \cite{Mosley2008NJP}. In the next section, we calculate the parameters for GVM conditions in detail \cite{Supplement2019}.

\begin{figure}[tbp]
\centering\includegraphics[width=8cm]{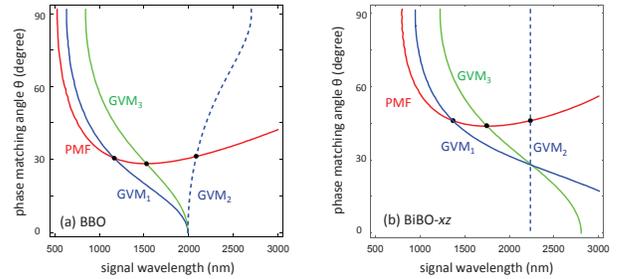}
\caption{The phase-matching function (PMF) and group-velocity matching functions (GVM$_1$, GVM$_2$, and GVM$_3$) for different signal wavelength $\lambda$ and phase-matching angle $\theta$ for BBO crystal (a) and for BiBO crystal in the $xz$ plane (b). In this calculation, we consider the Type-II  phase-matching condition with collinear and wavelength-degenerated ($2\lambda_p=\lambda_s=\lambda_i$) configuration.
 } \label{Fig:gvm3}
\end{figure}

\section{Calculation and Simulation}
\subsection{Wavelength degenerate case}

\begin{figure}[tbp]
\centering\includegraphics[width=9cm]{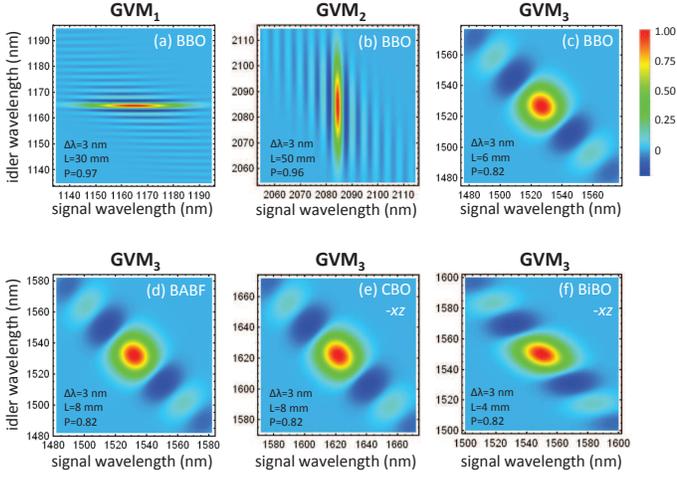}
\caption{ The JSA of the biphotons generated from BBO, BABF, CBO, and BiBO. The bandwidth  $\Delta \lambda$, the crystal length $L$ and the purity $P$ are listed in each figure.
 } \label{Fig:JSA6}
\end{figure}

First, we consider the Type-II ($e \to o + e$)  SPDC with collinear and wavelength-degenerated ($2\lambda_p=\lambda_s =\lambda_i $) configurations.
According to the three GVM conditions,  the GVM wavelength $\lambda_{p(s, i)}$ and the corresponding phase-matched angle $\theta$ can be calculated.
Take BBO as an example, Fig.\,\ref{Fig:gvm3}(a) shows the PMF and GVM$_{1(2,3)}$ conditions for different wavelengths and phase-matched angles. The cross points in Fig.\,\ref{Fig:gvm3} show that the PMF and GVM conditions are simultaneously satisfied at the wavelength of  1164 nm, 1526 nm, 2084 nm respectively.
Following the same method, we also calculate the GVM conditions for other crystals. Table\,\ref{table:uniaxial} lists the details of the three kinds of GVM conditions for 7 uniaxial crystals from the `BBO family'. In Tab.\,\ref{table:uniaxial}, the down-converted photons have a wavelength range from 1032 to 2158 nm, and the corresponding pump wavelength range is from 516 to 1079 nm.
It is noteworthy that BBO and BABF crystal can match the GVM$_3$ condition at 1526 nm and 1532 nm wavelength, e.g., in the C-band of telecommunication wavelength.
Figure\,\ref{Fig:JSA6}(a, b, c) show the JSAs of BBO crystal at its GVM$_{1(2,3)}$ wavelengths, and the spectral purity is 0.97, 0.96, and 0.82 respectively.
\begin{table*}[tbp]
\centering
\begin{tabular}{ c|c c c}
\hline \hline
Name&GVM$_1$  (purity $ \approx $ 0.97)& GVM$_2$  (purity $ \approx $ 0.97) & GVM$_3$  (purity $ \approx $ 0.82) \\
\hline
BBO& \tabincell{l}{ $\lambda_p$=582 nm, $\lambda_{s,i}$=1164 nm \\  $\theta$ =30.6$^\circ$,  $d_\textrm{eff}$=1.46 pm/V}
     & \tabincell{l}{ $\lambda_p$=1042 nm, $\lambda_{s,i}$=2084 nm \\  $\theta$ =31.2$^\circ$,  $d_\textrm{eff}$=1.32 pm/V}
      & \tabincell{l}{$\lambda_p$=763 nm, $\lambda_{s,i}$=1526 nm \\  $\theta$ =28.3$^\circ$,  $d_\textrm{eff}$=1.49 pm/V}  \\
\hline
CLBO& \tabincell{l}{$\lambda_p$=520 nm, $\lambda_{s,i}$=1040 nm \\  $\theta$ =43.0$^\circ$,  $d_\textrm{eff}$=0.68 pm/V}
      &  \tabincell{l}{$\lambda_p$=934 nm, $\lambda_{s,i}$=1868 nm \\  $\theta$ =44.2$^\circ$,  $d_\textrm{eff}$=0.62 pm/V}
      &  \tabincell{l}{$\lambda_p$=681 nm, $\lambda_{s,i}$=1362 nm \\  $\theta$ =39.4$^\circ$,  $d_\textrm{eff}$=0.65 pm/V}  \\

\hline
KABO& \tabincell{l}{$\lambda_p$=521 nm, $\lambda_{s,i}$=1042 nm \\  $\theta$ =40.8$^\circ$,  $d_\textrm{eff}$=0.24 pm/V}
      &  \tabincell{l}{$\lambda_p$=930 nm, $\lambda_{s,i}$=1860 nm \\  $\theta$ =42.0$^\circ$,  $d_\textrm{eff}$=0.23 pm/V}
      &  \tabincell{l}{$\lambda_p$=678 nm, $\lambda_{s,i}$=1356 nm \\  $\theta$ =37.6$^\circ$,  $d_\textrm{eff}$=0.27 pm/V}  \\
\hline
KBBF& \tabincell{l}{$\lambda_p$=516 nm, $\lambda_{s,i}$=1032 nm \\  $\theta$ =28.9$^\circ$,  $d_\textrm{eff}$=0.34 pm/V}
        &  \tabincell{l}{$\lambda_p$=933 nm, $\lambda_{s,i}$=1866 nm \\  $\theta$ =29.4$^\circ$,  $d_\textrm{eff}$=0.31 pm/V}
        &  \tabincell{l}{$\lambda_p$=682 nm, $\lambda_{s,i}$=1364 nm \\  $\theta$ =26.7$^\circ$,  $d_\textrm{eff}$=0.35 pm/V}  \\
\hline
 RBBF & \tabincell{l}{$\lambda_p$=533 nm, $\lambda_{s,i}$=1066 nm \\  $\theta$ =30.4$^\circ$,  $d_\textrm{eff}$=0.32 pm/V}
      &  \tabincell{l}{$\lambda_p$=974 nm, $\lambda_{s,i}$=1948 nm \\  $\theta$ =31.0$^\circ$,  $d_\textrm{eff}$=0.29 pm/V}
      &  \tabincell{l}{$\lambda_p$=708 nm, $\lambda_{s,i}$=1416 nm \\  $\theta$ =28.0$^\circ$,  $d_\textrm{eff}$=0.32 pm/V}  \\
\hline
 CBBF & \tabincell{l}{$\lambda_p$=529 nm, $\lambda_{s,i}$=1058 nm \\  $\theta$ =35.6$^\circ$,  $d_\textrm{eff}$=0.31 pm/V}
      &  \tabincell{l}{$\lambda_p$=940 nm, $\lambda_{s,i}$=1880 nm \\  $\theta$ =36.2$^\circ$,  $d_\textrm{eff}$=0.28 pm/V}
      &  \tabincell{l}{$\lambda_p$=694 nm, $\lambda_{s,i}$=1388nm \\  $\theta$ =32.8$^\circ$,  $d_\textrm{eff}$=0.32 pm/V}  \\
\hline
BABF*& \tabincell{l}{$\lambda_p$=578 nm, $\lambda_{s,i}$=1156nm \\  $\theta$ =47.6$^\circ$,  $d_\textrm{eff}$=0.53 pm/V}
      &  \tabincell{l}{$\lambda_p$=1079 nm, $\lambda_{s,i}$=2158 nm \\  $\theta$ =49.5$^\circ$,  $d_\textrm{eff}$=0.50 pm/V}
      &  \tabincell{l}{$\lambda_p$=766 nm, $\lambda_{s,i}$=1532 nm \\  $\theta$ =43.2$^\circ$,  $d_\textrm{eff}$= 0.63 pm/V}  \\
\hline \hline
\end{tabular}
\caption{\label{table:uniaxial}
Three kinds of GVM conditions for 7 uniaxial borate crystals. $\lambda_{p(s,i)}$ is the GVM wavelength for the pump (signal, idler). $\theta$ is the phase-matching angle and $d_\textrm{eff}$ is the effective nonlinear coefficient.
Most for the $d_\textrm{eff}$ values can be obtained from the SNLO $v70$ software package, developed by AS-Photonics, LLC \cite{SNLO70}.
*The $d_\textrm{eff}$ value  for BABF is not available from SNLO, so, we calculated the $d_\textrm{eff}$ using the method in the Appendix. The Sellmeier equations are obtained from Refs. \cite{Chen2012BOOK, Dmitriev1999, Nikogosyan2005}.
}
\end{table*}
\begin{table*}[tbp]
\centering
\begin{tabular}{ c|c c c}
\hline \hline
Name&GVM$_1$  (purity $ \approx $ 0.97)& GVM$_2$  (purity $ \approx $ 0.97) & GVM$_3$  (purity $ \approx $ 0.82) \\
\hline
\hline
BiBO$-{xz}$& \tabincell{l}{$\lambda_p$=687 nm, $\lambda_{s,i}$=1374 nm \\  $\theta$ =46.0$^\circ$,  $d_\textrm{eff}$=2.50 pm/V}
      &  \tabincell{l}{$\lambda_p$=1119 nm, $\lambda_{s,i}$=2238 nm \\  $\theta$ =46.0$^\circ$,  $d_\textrm{eff}$=2.30 pm/V}
      &  \tabincell{l}{$\lambda_p$=875 nm, $\lambda_{s,i}$=1750 nm \\  $\theta$ =43.8$^\circ$,  $d_\textrm{eff}$=2.48 pm/V}  \\
\hline
LBO$-{xz}$& not satisfied      & not satisfied
      & \tabincell{l}{$\lambda_p$=647 nm, $\lambda_{s,i}$=1294 nm \\  $\theta$ =4.7$^\circ$,  $d_\textrm{eff}$=-0.64 pm/V}  \\
\hline
CBO$-{xz}$& not satisfied      & not satisfied
      & \tabincell{l}{$\lambda_p$=811 nm, $\lambda_{s,i}$=1622 nm \\  $\theta$ =6.3$^\circ$,  $d_\textrm{eff}$=-0.21 pm/V}  \\
\hline \hline

\hline
LBO$-{yz}$& \tabincell{l}{ $\lambda_p$=491 nm, $\lambda_{s,i}$=982 nm \\  $\theta$ =31.8$^\circ$,  $d_\textrm{eff}$=-0.58 pm/V}
     & \tabincell{l}{ $\lambda_p$=864 nm, $\lambda_{s,i}$=1728 nm \\  $\theta$ =33.9$^\circ$,  $d_\textrm{eff}$=-0.52 pm/V}
      & not satisfied\\
\hline
LCB$-{yz}$& \tabincell{l}{ $\lambda_p$=537 nm, $\lambda_{s,i}$=1074 nm \\  $\theta$ =53.3$^\circ$,  $d_\textrm{eff}$=0.36 pm/V}
     & \tabincell{l}{ $\lambda_p$=944 nm, $\lambda_{s,i}$=1888 nm \\  $\theta$ =54.4$^\circ$,  $d_\textrm{eff}$=0.32 pm/V}
      & \tabincell{l}{$\lambda_p$=706 nm, $\lambda_{s,i}$=1412 nm \\  $\theta$ =47.9$^\circ$,  $d_\textrm{eff}$=0.39 pm/V}  \\
\hline
YCOB$-{yz}$& not satisfied
     & \tabincell{l}{ $\lambda_p$=1244 nm, $\lambda_{s,i}$=2488 nm \\  $\theta$ =17.1$^\circ$,  $d_\textrm{eff}$=0.14 pm/V}
      & not satisfied
      \\
\hline
GdCOB$-{yz}$& \tabincell{l}{ $\lambda_p$=668 nm, $\lambda_{s,i}$=1336 nm \\  $\theta$ =48.7$^\circ$,  $d_\textrm{eff}$=0.22 pm/V}
     & not satisfied
      & not satisfied
      \\
\hline \hline

\hline
CBO$-{xy}$& \tabincell{l}{$\lambda_p$=546 nm, $\lambda_{s,i}$=1092 nm \\  $\varphi$ =5.5$^\circ$,  $d_\textrm{eff}$=-0.23 pm/V}
     & \tabincell{l}{$\lambda_p$=1064 nm, $\lambda_{s,i}$=2128 nm \\  $\varphi$ =22.6$^\circ$,  $d_\textrm{eff}$=-0.73 pm/V}
      & not satisfied
        \\
\hline
LRB4$-{xy}$* &\tabincell{l}{$\lambda_p$=547 nm, $\lambda_{s,i}$=1148 nm \\  $\varphi$ =62.4$^\circ$,  $d_\textrm{eff}$=0.36 pm/V}
     &\tabincell{l}{$\lambda_p$=818 nm, $\lambda_{s,i}$=1636 nm \\  $\varphi$ =60.0$^\circ$,  $d_\textrm{eff}$=0.39 pm/V}
      &\tabincell{l}{$\lambda_p$=705 nm, $\lambda_{s,i}$=1410 nm \\  $\varphi$ =58.2$^\circ$,  $d_\textrm{eff}$=0.40 pm/V}
      \\
\hline
YCOB$-{xy}$& \tabincell{l}{$\lambda_p$=638 nm, $\lambda_{s,i}$=1276 nm \\  $\varphi$ =55.2$^\circ$,  $d_\textrm{eff}$=0.25 pm/V}
     & \tabincell{l}{$\lambda_p$=1205 nm, $\lambda_{s,i}$=2410 nm \\  $\varphi$ =60.5$^\circ$,  $d_\textrm{eff}$=0.06 pm/V}
      & \tabincell{l}{$\lambda_p$=842 nm, $\lambda_{s,i}$=1684 nm \\  $\varphi$ =47.1$^\circ$,  $d_\textrm{eff}$=0.46 pm/V}  \\
\hline
GdCOB$-{xy}$& \tabincell{l}{$\lambda_p$=668 nm, $\lambda_{s,i}$=1336 nm \\  $\varphi$ =68.8$^\circ$,  $d_\textrm{eff}$=-0.01 pm/V}
     & not satisfied
      & \tabincell{l}{$\lambda_p$=885 nm, $\lambda_{s,i}$=1770 nm \\  $\varphi$ =56.5$^\circ$,  $d_\textrm{eff}$=0.22 pm/V}  \\
\hline
\hline
\end{tabular}
\caption{\label{table:biaxial} Three kinds of GVM conditions for 7 biaxial borate crystals:
BiBO, LBO, and CBO with light propagating in the $xz$ plane; LBO, LCB, YCOB, and GdCOB  in the $yz$ plane; CBO, LRB4, YCOB, and GdCOB  in the $xy$ plane.
$\theta$ is the polar angle and $\varphi$ is the azimuth angle.
* The $d_\textrm{eff}$  for LRB4 is calculated using the method in the Appendix, and the $d_\textrm{eff}$  for other crystals are obtained from SNLO $v70$.
}
\end{table*}

For biaxial crystals, we consider the GVM conditions in different planes.
Figure\,\ref{Fig:gvm3}(b) shows the PMF and GVM$_{1(2,3)}$ conditions of BiBO crystals in the $xz$ plane for different wavelengths and phase-matched angles. The GVM$_{1(2,3)}$ wavelengths are at  1374 nm, 2238 nm, and 1750 nm respectively.
The GVM wavelengths for other crystals in different planes are listed in Tab.\,\ref{table:biaxial}.
It can be found that the BiBO, LBO, and CBO crystals can satisfy the conditions in the $xz$ plane, with $\varphi=0^\circ$.
LBO, LCB, YCOB, and GdCOB  can satisfy the conditions in the $yz$ plane, with $\varphi=90^\circ$.
CBO, LRB4, YCOB, and GdCOB can satisfy the conditions in the $xy$ plane, with $\theta=90^\circ$.
In Tab.\,\ref{table:biaxial}, the down-converted photons have a wavelength range from 982 to 2488 nm, and the corresponding pump wavelength range is from 491 to 1244 nm.
It is noteworthy that CBO in the $xz$ plane and LRB4  in the $xy$ plane can match the GVM$_3$  and GVM$_2$ conditions at 1622 nm and 1636 nm wavelength, i.e., in the L-band of telecommunication wavelength.

Figure\,\ref{Fig:JSA6}(d) shows the JSA of BABF crystal at its GVM$_3$ wavelength of 1532 nm with a  purity  of 0.82.
Figure\,\ref{Fig:JSA6}(e) shows  the JSA of CBO  at its GVM$_3$ wavelength in $xz$ plane, with the spectral purity of 0.82.
Under the GVM$_3$ condition, the purity can be kept over 0.8  when the wavelength is tuned over 200 nm wavelength range \cite{Jin2013OE}. Take BiBO crystal as an example, although the GVM$_3$ wavelength is 1750 nm for BiBO, the purity still can achieve 0.82 at 1550 nm, as shown in Fig\,\ref{Fig:JSA6}(f).
This property is very useful to increase the tunability of the pure single-photon source.

\subsection{Wavelength nondegenerate case}
\begin{figure}[tbp]
\centering\includegraphics[width=9cm]{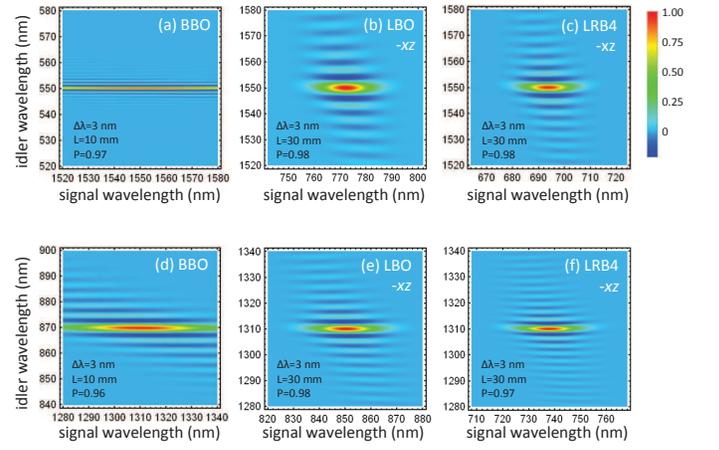}
\caption{ The JSA of biphotons with non-degenerated wavelengths generated from BBO (a, d), LBO (b, e) and LRB4 (c, f) under the GVM$_1$ condition. The bandwidth  $\Delta \lambda$, the crystal length $L$ and the purity $P$ are listed in each figure. Figures in the first (second) row are for 1550 (1310) nm wavelength.
 } \label{Fig:wavelnon}
\end{figure}

In the wavelength nondegenerate case, we consider the wavelength of the pump photons locate at 400 $\sim$ 540 nm, which are the wavelengths for commercial-available, low-cost laser diodes. One of the down-converted photons is at 500 $\sim$ 900 nm wavelength range, where silicon avalanche photodiode (APD) based single-photon detector has good performance. The other  down-converted  photon is at telecom wavelength around 1550 nm or 1310 nm, for low-loss long-distance transmission in optical fibers. The typical configurations include 405 nm $\to $ 548 nm + 1550 nm, 520 nm $\to $ 783 nm + 1550 nm,  405 nm $\to $ 586 nm + 1310 nm, and 520 nm $\to $ 862 nm + 1310 nm. These configurations are connecting the visible, near-infrared wavelength and the telecom wavelength \cite{Kaneda2016, Saglamyurek2011nature,  Zhou2016NC}.
In these wavelength range, our results show that 7 uniaxial and 2 biaxial crystals satisfy the GVM$_1$ condition in SPDC with collinear and wavelength-nondegenerated ($\lambda_s  \ne \lambda_i$) configurations.
Table \,\ref{table2} lists the GVM$_1$ condition for BBO, CLBO, KABO, KBBF, RBBF, CBBF, BABF, LBO, and LRB4.
The pump wavelength is ranging from 401 nm to 542 nm, and down-converted photons' wavelength includes the telecom wavelength of 1310 nm or 1550 nm.
The JSAs under these conditions for BBO, LBO, and LRB4 are shown in Fig.\,\ref{Fig:wavelnon}.
Each crystal can achieve the GVM$_1$ condition at both 1310 nm and 1550 nm.
All the JSAs in Fig.\,\ref{Fig:wavelnon} have a long elliptical distribution in the horizontal direction, and with high purities from 0.96 to 0.98.

\begin{table*}[tbp]
\centering
\begin{tabular}{ c|l l l}
\hline \hline
Name&GVM$_1$ condition (purity $ \approx$ 0.97) \\
\hline
BBO&  $\lambda_p$=406 nm, $\lambda_s$=1550 nm,  $\lambda_i$=550 nm,  $\theta$ =55.2$^\circ$,  $d_\textrm{eff}$=0.60 pm/V \\
\hline
BBO&    $\lambda_p$=523 nm, $\lambda_s$=1310 nm,  $\lambda_i$=870 nm,  $\theta$ =36.8$^\circ$,  $d_\textrm{eff}$=1.24 pm/V  \\
\hline
CLBO&  $\lambda_p$=401 nm, $\lambda_s$=1310 nm,  $\lambda_i$=579 nm,  $\theta$ =77.9$^\circ$,  $d_\textrm{eff}$=0.27 pm/V \\
\hline
KABO&  $\lambda_p$=403 nm, $\lambda_s$=1310 nm,  $\lambda_i$=583 nm,  $\theta$=68.4$^\circ$,  $d_\textrm{eff}$=0.07 pm/V \\
\hline
KBBF&  $\lambda_p$=542 nm, $\lambda_s$=925 nm,  $\lambda_i$=1310 nm,  $\theta$ =25.7$^\circ$,  $d_\textrm{eff}$=0.36 pm/V \\
\hline
RBBF&  $\lambda_p$=417 nm, $\lambda_s$=1310 nm,  $\lambda_i$=611 nm,  $\theta$ =45.0$^\circ$,  $d_\textrm{eff}$=0.21 pm/V \\
\hline
CBBF&  $\lambda_p$=414 nm, $\lambda_s$=1310 nm,  $\lambda_i$=605 nm,  $\theta$ =54.4$^\circ$,  $d_\textrm{eff}$=0.16 pm/V \\
\hline
BABF&  $\lambda_p$=509 nm, $\lambda_s$=1310 nm,  $\lambda_i$=833 nm,  $\theta$ =62.8$^\circ$,  $d_\textrm{eff}$=0.24 pm/V*  \\
\hline
LBO-xz&  $\lambda_p$=515 nm, $\lambda_s$=772 nm,  $\lambda_i$=1550 nm,  $\theta$ =22.0$^\circ$,  $d_\textrm{eff}$=-0.47 pm/V  \\
\hline
LBO-xz&  $\lambda_p$=516 nm, $\lambda_s$=850 nm,  $\lambda_i$=1310 nm,  $\theta$ =16.0$^\circ$,  $d_\textrm{eff}$=-0.56 pm/V \\
\hline
LRB4-xz &  $\lambda_p$=479 nm, $\lambda_s$=694 nm,  $\lambda_i$=1550 nm,  $\theta$ =20.5$^\circ$,  $d_\textrm{eff}$=-0.31 pm/V* \\
\hline
LRB4-xz &  $\lambda_p$=472 nm, $\lambda_s$=738 nm,  $\lambda_i$=1310 nm,  $\theta$ =7.1$^\circ$,  $d_\textrm{eff}$=-0.12 pm/V* \\
\hline \hline
\end{tabular}
\caption{\label{table2} The parameters of wavelength non-degenerated SPDC for 7 uniaxial crystals and 2 biaxial crystals. $\theta$ is the phase-matching angle and $d_\textrm{eff}$ is the effective nonlinear coefficient.
* The $d_\textrm{eff}$ values for BABF and LRB4 are not available in SNLO $v70$, and are calculated using the method shown in the Appendix. The $d_\textrm{eff}$ values for other crystals are taken from the SNLO $v70$ software package.
 }
\end{table*}

\subsection{HOM interference measurement}
The quality of the spectrally uncorrelated biphoton state can be tested by  Hong-Ou-Mandel (HOM)  interference. There are two kinds of HOM interference, the first one is the HOM interferences using signal and idler photons from the same SPDC source, with a typical setup shown in \cite{Hong1987} . In this case, the two-fold coincidence probability $ P_2(\tau )$ as a function of the time delay $\tau$  is given by \cite{Grice1997, Ou2007, Jin2015OE}:
\begin{equation}\label{eq:P2}
\begin{split}
P_2(\tau ) =  & \frac{1}{4} \int\limits_0^\infty  \int\limits_0^\infty  d\omega _s  d\omega _i \\
              & \left| {[f(\omega _s ,\omega _i ) - f(\omega _i ,\omega _s )e^{ - i(\omega _s  - \omega _i )\tau } ]} \right|^2.
\end{split}
\end{equation}
The second one is the HOM interference  with two independent heralded single-photon sources, with a typical experimental setup shown in Refs. \cite{Mosley2008PRL, Jin2013PRA}.
In this interference, two signals $s_{1}$ and $s_{2}$ are sent to a beamsplitter for interference, and two idlers $i_{1}$ and $i_{2}$ are detected by single-photon detectors for heralding the signals.
The four-fold coincidence counts $P_4$ as a function of $\tau$ can be described by \cite{Ou2007, Jin2015OE}
\begin{equation}\label{eq:P4}
\begin{split}
P_4 (\tau )  = & \frac{1}{4}  \int_0^\infty \int_0^\infty \int_0^\infty \int_0^\infty d\omega _{s_1} d\omega _{s_2} d\omega _{i_1} d\omega _{i_2}  \\ & {\rm{|}}f_1 (\omega _{s_1} ,\omega _{i_1} )f_2 (\omega _{s_2} ,\omega _{i_2} )- \\ & f_1 (\omega _{s_2} ,\omega _{i_1} )f_2 (\omega _{s_1} ,\omega _{i_2} )e^{ - i(\omega _{s_2}  - \omega _{s_1} )\tau } {\rm{|}}^{\rm{2}},
\end{split}
\end{equation}
where $f_1$ and $f_2$ are the JSAs from the first and the second crystals.

\begin{figure}[tbp]
\centering\includegraphics[width=8cm]{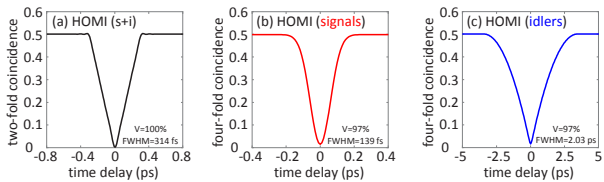}
\caption{(a) HOM interference curve with a signal and idler from the same BBO crystal, with the JSA shown in Fig.\,\ref{Fig:JSA6}(c). (b) and (c) show the HOM interference curve with two heralded signal or idler photons from two independent BBO sources, with the  JSA shown in Fig.\,\ref{Fig:JSA6}(a). The visibility (V) and full-width at half-maximum (FWHM) values are labeled in each figure.
 } \label{Fig:homi}
\end{figure}
Figure\,\ref{Fig:homi}(a) is the obtained HOM interference pattern between a signal and an idler from the same BBO crystal, with the JSA shown in Fig.\,\ref{Fig:JSA6}(b). Under this condition, the visibility is as high as 100\%.
In this case, the HOM interference visibility is determined by the exchanging symmetry of the biphotons, i.e., $f(\omega_s,\omega_i)=f(\omega_i,\omega_s)$.
Figure\,\ref{Fig:homi}(b, c) are the HOM interference curves between two heralded signals or two heralded idlers. Without using any narrow bandpass filters, visibility can achieve 98\%. Please note that the visibility equals to the purity for the ideal case in the HOM interference between independent sources \cite{Mosley2008PRL, Takesue2012, Jin2013PRA}.

\section{Discussion}

In this study, we considered the GVM conditions in the main planes for biaxial crystals, i.e., in the $xz$, $yz$ and $xy$ planes,  it is also possible to realize GVM conditions in an arbitrary direction. For example, previous studies have investigated phase-matching conditions and the effective nonlinear coefficient in BiBO crystal in an arbitrary direction \cite{Halevy2011, Chenluokan2017}.
Further, we have considered the GVM condition in the collinear condition, it is also possible to realize pure-state generation in the non-collinear condition, since non-collinear configurations have been widely used in entangled photon source design \cite{Kwiat1995, Kwiat1999}.
In addition, the  `BBO family'  has  a lot of borate crystals \cite{Chen2012BOOK}, and we only considered 14 crystals in this work, because the Sellmeier equations for many borate crystals are still not available.
Therefore, to investigate the generation of  spectrally uncorrelated biphotons from more borate crystals, in an arbitrary direction, in the nonlinear configuration for higher nonlinear coefficient and better wavelength range will be the future work. Especially, it is promising to study GVM conditions in BiBO crystal in the in an arbitrary direction, since this crystal has the highest nonlinear coefficient in Table\,\ref{table:CrystalSummary}.

Designing new borate crystals with novel characteristics (e.g., higher nonlinear coefficient, smaller walk-off effect, better wavelength range) is still an active research area in the study of nonlinear optical crystals \cite{Yang2017crystal, Kang2018RRL, Zhang2019asc}.
For example, following the anionic group theory, new nonlinear optical   materials can be designed through the combination of borate and other material genomes \cite{Kang2018RRL, Zhang2019asc}.
In the future, more and more  borate crystals might be synthesized.  Therefore, it is also possible to design and discover more borate crystals to prepare spectrally pure single-photon sources.

From the perspective of future applications,  the biphotons in Fig.\,\ref{Fig:JSA6}(c-f) and Fig.\,\ref{Fig:wavelnon} can be applied for  telecom wavelength;  the biphotons in Fig.\,\ref{Fig:wavelnon}  are good candidates for connecting the  visible-near-infrared  wavelength and telecom wavelength, which has great potential in quantum networks~\cite{Sangouard2011rmp}.
Crystals with GVM$_1$ and GVM$_2$ conditions can be used for interference between two signals from two independent SPDC sources. Crystals with GVM$_3$ condition are useful for interference between the signal and idler photons from one SPDC source.
The highly pure single-photon sources at telecom wavelengths are very important for practical applications that require long-distance transmission in low-loss and low-cost optical fibers.

\section{Conclusion}
In summary, we have theoretically investigated the preparation of spectrally uncorrelated biphoton state and heralded pure single-photon state from 14 borate crystals. It was  shown that these 14  crystals from the `BBO family', namely BBO, CLBO, KABO, KBBF, RBBF, CBBF, BABF,  BiBO, LBO, CBO, LRB4, LCB, YCOB and GdCOB can satisfy three kinds of GVM condition from near-infrared to telecom bands. These crystals can be used to prepare spectrally uncorrelated biphoton state with purity as high as 0.98. Visibility of 100\%  in the HOM interference with signal and idler photons from one SPDC source, and visibility of 98\% in the HOM interference between two independent SPDC sources can be achieved. This work will provide good single-photon sources for photonic quantum information from near-infrared to telecom wavelength, and will inspire people to find more crystal materials with superior properties for specific applications.

\section*{Acknowledgments}
This work is  partially supported by the National Key R$\&$D Program of China (Grant No. 2018YFA0307400), the National Natural Science Foundations of China (Grant Nos.91836102, 11704290, 61775025, 61704164, 61405030), and by the Program of State Key Laboratory of Quantum Optics and Quantum Optics Devices (No: KF201813).

\section*{Appendix}
\subsection*{A1: Coordinate for uniaxial and biaxial crystals in the calculation }
\begin{figure}[!htbp]
\centering\includegraphics[width=8cm]{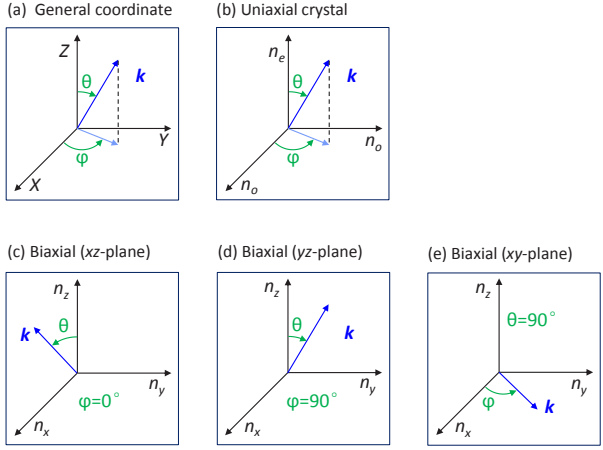}
\caption{The refractive index coordinate for uniaxial and biaxial crystals in the calculation. (a) is a general coordinate; (b) is the refractive index coordinate for uniaxial crystals. (c-e) are refractive index coordinate for biaxial crystals in the $xz$, $yz$ and $xy$ planes.
 } \label{A1coordinate}
\end{figure}

Figure\,\ref{A1coordinate} shows the refractive index coordinate for uniaxial and biaxial crystals in the calculation.
$\theta$ is the polar angle and $\varphi$ is the azimuth angle.

\subsection*{A2: Calculation of effective nonlinear coefficient }
In this section, we discuss how to calculate the effective nonlinear coefficient ($d_{eff}$ ) for uniaxial and biaxial crystals.
Firstly, we consider the uniaxial crystals  and  take BBO crystal as an example.

The $d_{eff}$  value is determined by point group and phase-matching condition.
For BBO crystal with a point group 3m \cite{Nikogosyan1991}, under the Type-II phase-matching  condition,  the $d_{eff}$ could be calculated using the following equation \cite{Midwinter1965, Dmitriev1999}:
\begin{equation}\label{a1}
  d_{eff}(\text{BBO}) = d_{22}\cos^{2}(\theta + \rho)\cos(3\varphi).
\end{equation}
In order to maximize $d_{eff}$, the item $\cos(3\varphi)$ should reach a maximum value, i.e, $\cos(3\varphi)=1$ for $\varphi=0^\circ$ or  $120^\circ$.
$\rho$ is the walk-off angle, which can be expressed as \cite{Dmitriev1999}:
\begin{equation}\label{a2}
  \rho(\theta) = \pm \arctan[\frac{n^2_o(\lambda)}{n^2_e(\lambda)}\tan(\theta)]\mp \theta,
\end{equation}
where the upper (lower) signs are for negative (positive) crystals. For BBO, a negative crystal, the equation is:
\begin{equation}\label{a3}
  \rho(\theta)(\text{BBO}) = \arctan[\frac{n^2_o(\lambda)}{n^2_e(\lambda)}\tan(\theta)] - \theta.
\end{equation}
Finally, we can obtain $d_{eff}$ by substituting the value of $d_{22}$ = 2.2 pm/V \cite{Shoji1999}, phase-matching angle, and wavelength into the above equation.

For uniaxial crystals, we take LBO and BiBO as two examples to calculate the $d_{eff}$.
LBO is a negative biaxial crystal with the point group of mm2,  $d_{31}$ = $-$ 0.67 pm/V, and $d_{32}$ = 0.85 pm/V \cite{Roberts1992}.
Under the type-II phase-matching condition, the expression of $d_{eff}$ on the $yz$ plane and $xz$ plane can be calculated as follow \cite{Dmitriev1993, Dmitriev1999}:

$xz-$plane:
\begin{equation}\label{a4}
  d_{eff(\text{LBO}-xz)} =d_{32}\sin^2(\theta + \rho) + d_{31}\cos^2(\theta + \rho).
\end{equation}

$yz-$ plane:
\begin{equation}\label{a5}
  d_{eff( \text{LBO}-yz   )}= d_{31}\cos(\theta + \rho).
\end{equation}

For BiBO crystal,  a negative biaxial crystal with the point group of 2 and $d_{26}$ =3.48 pm/V, the $d_{eff}$ on  the $xz$ plane is determined by \cite{Chen2012BOOK}:
\begin{equation}\label{a6}
  d_{eff(\text{BiBO}-xz)} =d_{26}\cos (\theta + \rho).
\end{equation}

\end{document}